\documentclass[twocolumn,showpacs,preprintnumbers,amsmath,amssymb,prl,superscriptaddress,nofootinbib]{revtex4}
\usepackage{graphicx} 
\usepackage{dcolumn} 
\usepackage{bm} 
\usepackage{amssymb} 
\usepackage{amsmath} 
\usepackage{amsfonts}
\usepackage{color}

\definecolor{darkgreen}{RGB}{50,150,0}

\begin{document}

\title{Fine Tuning, Sequestering, and the Swampland}

\author{Jonathan J. Heckman}
\affiliation{Department of Physics and Astronomy, University of Pennsylvania, Philadelphia, PA 19104, USA}
\author{Cumrun Vafa}
\affiliation{Jefferson Physical Laboratory, Harvard University, Cambridge, MA 02138, USA}

\begin{abstract}

We conjecture and present evidence that any effective field theory coupled to gravity in flat space admits at most a finite number of fine tunings, depending on the amount of supersymmetry and spacetime dimension. In particular, this means that there are infinitely many non-trivial correlations between the allowed deformations of a given effective field theory in the gravitational context.
Fine tuning of parameters allows us to obtain some consistent CFTs in the IR limit of gravitational theories.   Related to finiteness of fine tunings, we conjecture that except for a finite number of CFTs, the rest cannot be consistently coupled to gravity and belong to the swampland.
Moreover, we argue that even though matter sectors coupled to gravity may sometimes be partially sequestered, there is an irreducible level of
mixing between them, correlating and coupling infinitely many operators between these sectors.

\end{abstract}

\maketitle

\subsection{Introduction}
\label{sec:int}

The standard lore of effective field theory organizes
physics according to energy scales.
Once the ultraviolet (UV) boundary conditions of
an effective field theory are specified,
the consequences for low energy physics in principle follow
from an analysis of renormalization group (RG) flow. It is natural to
ask whether the parameters of the UV completion
are completely arbitrary ``order one'' constants, or whether
additional structure is required for consistency with quantum gravity. This
has direct bearing on a number of hierarchy problems in particle physics, including the
radiative stability of the Higgs mass relative to the Planck scale and the impact of
extra sectors on the Standard Models of particle physics and cosmology.

In this note we argue that for a given effective field theory in flat space
coupled to quantum gravity in the same spacetime dimension,\footnote{Note that a stack of D3-branes
in 10D flat space would not satisfy this condition.}
\textit{only a finite number of parameters can ever be tuned}.
More precisely, in an effective field theory with cutoff
the Planck scale $M_{pl}$ and Lagrangian:
\begin{equation}
L_{\text{eff}} = \sum_{i} \frac{g_{i}}{M_{pl}^{\nu_{i}}} \mathcal{O}_{i},
\end{equation}
we argue that the infinite set of couplings $g_{i}$
are all determined by a \textit{finite} list of parameters
in the effective field theory. Moreover, we conjecture this number only
depends on the number of supersymmetries and the macroscopic
dimension of spacetime.

The primary evidence for our conjecture is that there appear to be
a finite number of Minkowski vacua arising in string theory compactifications. For example, it
is widely believed that there are a finite number of Calabi-Yau threefolds, and tadpole cancellation
conditions enforce sharp upper bounds on the number of flux quanta / discrete parameters.
These statements have natural generalizations in the context of
M- and F-theory compactifications as well. For related discussion
of finiteness in the string theory landscape, see reference \cite{Acharya:2006zw}.

Even so, the consequences for low energy physics of these simple observations
appear to have not been completely spelled out. Finiteness in the number of compactification
geometries means there are also at most a finite number of tunable parameters in any given effective field theory.
This greatly limits the structure of low energy effective field theories which can emerge in the infrared (IR)
of a consistent string compactification. For example, assuming that there are only a finite number of
conformal field theories (CFTs) for a fixed total number of relevant deformations, it implies there are a
finite number of conformal fixed points which can be realized in quantum gravity. Additionally,
distinct field theory sectors which are decoupled in the $M_{pl} \rightarrow \infty$ limit
are necessarily always coupled when gravity is not switched off. While this latter statement may not at
first appear surprising, finiteness in the number of allowed compactifications and their moduli means that there is
an infinite amount of correlated mixing between extra sectors.

The plan in the rest of this note will be to expand on the above points. First, we
discuss finiteness of the number of CFTs which can consistently couple to gravity, also illustrating
with examples in different dimensions and amounts of supersymmetry. We then discuss some
additional phenomenological consequences of having a finite number of tunable couplings.

\subsection{CFTs and Quantum Gravity}

In this section we consider the coupling of a conformal field theory (CFT)
in flat space to gravity. CFTs are central to many aspects of quantum field theory (QFT),
because most QFTs can be viewed as moving away from, or towards a conformal fixed point.
More precisely, by ``coupling a CFT to gravity'' we mean there is a
gravitational theory with an IR limit which includes the corresponding CFT.

The passage towards and away from a fixed point is dictated by its space of possible
deformations. In general terms, such deformations can arise either
through varying the couplings of the CFT or through operator VEVs.
We refer to these as the ``parameters'' of the CFT.

If we consider the coupling of such a theory to gravity, the parameters
controlling such deformations will correspond to dynamical fields,
as in general any quantum gravitational theory is conjectured to have no free parameters
\cite{Ooguri:2006in}. Indeed, in the context of string
compactification, one should expect that \textit{all} of the parameters
appearing in a putative CFT originate from vevs of fields which are light relative to the Planck scale.
Such fields could originate from operators within the CFT itself, or from fields outside of the CFT coupled to gravity.
But such parameters cannot be tuned arbitrarily. For quantum gravity in Minkowski space,
this follows directly from the conjectured finiteness of such vacua in string theory
\cite{Ooguri:2006in}.\footnote{For AdS vacua \cite{LPV} there is
a refined version of this conjecture which appears to hold. The reason
for the caveat is that even though fluxes allow one to tune the size of the AdS space, this is
correlated with the size of the extra dimensions.} Combining these
statements, we learn that of the a priori infinitely
many possible deformations of a CFT, only a \textit{finite}
number of them can be fine tuned, and the rest are determined in
terms of the fine tuned ones!

 This in particular implies that if we
want to obtain a CFT in the infrared limit of a gravitational theory, it had
better be the case that the total number of relevant deformations
and VEVs that takes us away from the conformal point are not too large.  Otherwise, we may not have enough light modes in the gravity theory to fine tune in order to realize the CFT point.  Assuming that for a fixed number of relevant deformations there are only a finite number of CFTs, which seems quite reasonable, we thus conclude that  only a finite number of CFTs can be
realized as IR limits of a gravitational theory.  In other words {\it nearly all CFTs
belong to the swampland! }

Of course it is not a priori clear that we can get {\it any } CFT in the infrared limit of a
gravitational theory.  As we shortly explain, however, these are abundant
in string constructions, and arise by tuning all relevant
deformations to zero in a given string compactification. Moreover, we find that this fine tuning is not limited only to relevant deformations and in some cases
some of the irrelevant operators can also be tuned to zero.
But we will find that
almost all irrelevant parameters are necessarily switched on in quantum gravity.

Having spelled out the general reason to expect such conjectures to be true, we now
turn to explicit examples illustrating the main points.

\subsection{Examples}

We organize our discussion of examples according
to the number of spacetime dimensions, as well as the number of
supersymmetries.


Consider first the case of 6D SCFTs. Such theories admit either
$\mathcal{N} = (2,0)$ or $\mathcal{N} = (1,0)$ supersymmetry,
namely sixteen or eight real supercharges, respectively. In both cases, there is a
conjectural classification of the resulting theories which can result from
string / F-theory compactification. An interesting feature of these examples
is that there are no supersymmetric relevant or marginal deformations of these theories \cite{Louis:2015mka, Cordova:2016xhm}.
Instead, all the parameters are obtained from operator VEVs.

For interacting 6D SCFTs with $(2,0)$ supersymmetry there is a famous ADE classification which is obtained from
compactifications of type IIB on hyperkahler geometries with an ADE singularity \cite{Witten:1995zh}. Such geometries
all have the local presentation $\mathbb{C}^2 / \Gamma_{ADE}$ with $\Gamma_{ADE} \subset SU(2)$ a finite order subgroup.\footnote{There is
also some choice in the topological data of these theories associated with the spectrum of various defect operators
(see e.g. \cite{Gaiotto:2014kfa}). This will not play an important role in our considerations.}
All other $(2,0)$ SCFTs are then obtained by taking tensor products of these
basic building blocks. Such theories are then classified by semi-simple Lie algebras $\mathfrak{g}_1 \times ... \times \mathfrak{g}_n$ with
each $\mathfrak{g}_i$ a simple ADE Lie algebra. In all these cases, the space of possible deformations is controlled by the number of
tensor multiplets which is given by the rank of the corresponding semi-simple Lie algebra. This leads to an
infinite list of such SCFTs.

Even so, nearly all of these theories belong to the 6D swampland since they cannot be coupled to 6D $(2,0)$ supergravity.
In this case, the total rank of the semi-simple Lie algebra $\mathfrak{g}_1 \times ... \times \mathfrak{g}_n$ is bounded
above by:
\begin{equation}\label{6Drankbound}
\text{rank}(\mathfrak{g}_1 \times ... \times \mathfrak{g}_n) \leq 21
\end{equation}
This follows from anomaly cancellation for $(2,0)$ supergravity coupled to matter fields, which in this case fixes the total number of tensor multiplets to be $21$ (see reference \cite{AlvarezGaume:1983ig}). Thus, any non-trivial CFTs that can emerge must fit within this constraint. This is also consistent with the fact that to reach a $(2,0)$ supergravity theory from compactification of type IIB strings in the first place, we need to use a K3 surface, and there is only one manifold of this topology.

In fact, one can in principle also classify all possible 6D SCFTs which can be obtained from singular limits in the moduli space of IIB compactified on a K3 surface (see e.g. \cite{Witten:1995zh, Aspinwall:1996mn}):\footnote{Observe also that
compactifying on a further $S^1$ leads to a dual description in terms of heterotic strings on $T^5$.}
\begin{equation}
\mathcal{M}_{IIB} \simeq O(\Gamma_{5,21} ) \backslash O(5,21) / (O(5) \times O(21)).
\end{equation}
To get a 6D SCFT, we have the necessary condition that the root system for
the corresponding semi-simple Lie algebra embeds in $\Gamma_{5,21}$.
As two examples in this class, note that we can get a single group corresponding to the $D_{21}$ theory. As another example,
note that we can get three decoupled SCFTs, $E_8\times E_8\times A_5$ coupled
to $(2,0)$ supergravity in six dimensions. Both examples saturate the upper bound of $21$
tensor multiplets, and the latter example also shows we can have multiple decoupled CFTs in the deep infrared.

Consider next 6D theories with minimal $\mathcal{N} = (1,0)$ supersymmetry.
Anomaly cancellation, along with physical constraints such
as the requirement that all kinetic energy terms remain positive definite
imposes stringent conditions on admissible low energy theories.
Embedding such theories in string theory imposes additional limitations, cutting this to a finite list of
possibilities \cite{Kumar:2010ru}. The broadest
class of UV completions is obtained via compactifications
of F-theory (see e.g. \cite{Kumar:2010ru, Morrison:2012js, Morrison:2012np}
as well as the review \cite{Taylor:2011wt}). In F-theory, all such vacua
are realized by a choice of an elliptically fibered Calabi-Yau threefold
with base $B$ a K\"ahler surface. The dimension of the tensor branch moduli
space is $h^{1,1}(B) - 1$ while the number of hypermultiplets of the
Higgs branch moduli space is counted by $h^{2,1}$ of the Calabi-Yau threefold.
6D SCFT sectors arise from simultaneously collapsing $\mathbb{P}^1$'s in the base $B$,
and a classification of the infinite list of possibilities was completed
in references \cite{Heckman:2013pva, Heckman:2015bfa}
(see \cite{Heckman:2018jxk} for a review).
It is also possible to couple some of these 6D CFTs to gravity
(see e.g. \cite{DelZotto:2014fia, Anderson:2018heq, Hayashi:2019fsa}).
An example of this sort is to take the Calabi-Yau threefold $(T^2 \times T^2 \times T^2) / \mathbb{Z}_3$, where we tune the complex
structure of each $T^2$ to admit a $\mathbb{Z}_3$ symmetry. We will revisit this example
in lower dimensions as obtained by compactification on further circles.

Nevertheless, almost all of the 6D $(1,0)$ CFTs, with the exception of a finite number of them,
require a non-compact Calabi-Yau threefold. This means they
cannot be dynamically coupled to gravity in six dimensions.
Let us briefly review this finiteness property.
There are general finiteness results on the existence of \textit{elliptically fibered}
Calabi-Yau threefolds \cite{Grassi, Gross} which in
turn imply upper bounds on the number of tensor, vector and hyper multiplets
\cite{Kumar:2010ru}. While the exact upper bound is unknown, in practice, the largest dimension for the tensor branch moduli space is
$193$, as found in \cite{Candelas:1997eh, Aspinwall:1997ye}, which fits with general statements available for F-theory models
with a toric base \cite{Morrison:2012js}. Indeed, there is a rigorous upper bound on
the Hodge numbers $h^{1,1}$ and $h^{2,1}$ of elliptic threefolds with a toric base \cite{Taylor:2012dr}, and there is a general
expectation that moving beyond the toric case will not greatly affect these bounds.
For our present purposes, this of course means that the number of 6D SCFTs which can be consistently coupled to 6D supergravity
is also finite.
Despite this finiteness in the allowed realizations of $(1,0)$ SCFTs
coupled to gravity arising in string theory, anomaly cancellation constraints allow
in principle infinitely many possible SCFTs coupling to gravity \cite{Kumar:2010ru}.
However, at least some of these infinite families can be ruled out by additional
consistency conditions \cite{KSV}, and it is natural to conjecture that when
all constraints are imposed only a finite number of them will survive.


In the case of 5D SCFTs, the superconformal algebra allows for $\mathcal{N} = 1$ supersymmetry, namely eight real
supercharges. One way to generate a large class of examples is to consider M-theory compactified on a Calabi-Yau threefold
with at least one divisor which collapses to a point at finite distance in moduli space.  Such
singularities are expected to generate most, if not all of the possible 5D SCFTs. The general point
is that these canonical singularities are localized at isolated patches of the Calabi-Yau, but are coupled to one another by effects
inherited from 11D supergravity.

One way to generate a large class of examples of this sort is to take a 6D $(1,0)$ SCFT and compactify it
on a circle \cite{DelZotto:2017pti,Jefferson:2017ahm}. An illustrative example of this type is given by the Calabi-Yau threefold $(T^2 \times T^2 \times T^2) / \mathbb{Z}_3$, where we tune the complex structure of each $T^2$ to admit a $\mathbb{Z}_3$ symmetry. There are precisely $27$ orbifold fixed points, each of which is locally characterized by the geometry $\mathbb{C}^{3} / \mathbb{Z}_3$ which in the resolved phase is described by a $\mathbb{P}^2$ collapsing to zero size. In the limit where the Calabi-Yau threefold decompactifies, we have $27$ 5D SCFTs decoupled from 5D supergravity, each of which is described by a
5D SCFT \cite{Morrison:1996xf,Douglas:1996xp}. The Coulomb branch is given by $27$ 5D $\mathcal{N} = 1$ vector multiplets, each with a real scalar.
The Coulomb branch parameter dictates the size of the $\mathbb{P}^2$, and there are BPS states obtained from M2-branes wrapped on curves in each $\mathbb{P}^2$. In particular, even in the limit where we couple to quantum gravity,
all $27$ Coulomb branch parameters can be tuned to zero independently.

But it is also true that when gravity is switched on, these different sectors cannot be completely decoupled.  To see this, suppose we go away from the CFT point by moving onto the Coulomb branch, giving finite size to each $\mathbb{P}^2$.  In this case, some of the BPS states which were previously massless at the CFT point (corresponding to M2-branes wrapped on curves inside each $\mathbb{P}^2$) now acquire a mass proportional to $m_i\sim \phi_i$ the Coulomb branch parameter. Here, we have normalized the scalar fields to have mass dimension one and fermions to have dimension two. If we denote by $L$ the characteristic length scale of the $T^2$'s, there is a Kaluza-Klein (KK) mass scale $M_{KK} \sim 1 / L$. Integrating out the whole tower of KK states from compactification of the 11D supergravity model will induce various higher-dimension operators in the 5D effective field theory. This includes higher derivative interactions as well as four-fermion interactions. Indeed, letting $\psi_i$ denote one such fermionic field describing 5D excitations of an M2-brane wrapped on a curve, it is not difficult to see that exchange of the KK tower of 11D supergravity modes induces interactions between fermions in previously decoupled sectors of the form:\footnote{One way to derive this coupling is to consider the force between wrapped M2-branes and anti-M2-branes.}
\begin{equation}
L_{mix} \supset {m_im_j\over M_{pl}^3 M_{KK}^2}{\overline\psi_i}\psi_i  {\overline \psi_j }\psi_j,
\end{equation}
with $M_{pl}$ the 5D Planck scale. Since we also have $m_i \sim \phi_i$, we
learn that there is a mixing term of the form:
\begin{equation}
L_{mix} \supset C_{ij}{\cal O}_i{\cal O}_j
\end{equation}
where ${\cal O}_i=\phi_i{\overline \psi_i}\psi_i , \quad {\cal O}_j=\phi_j{\overline \psi_j}\psi_j$, and
\begin{equation}\label{ccoeff}
C_{ij}\sim {1\over M_{pl}^5} \left({M_{pl}\over M_{KK}} \right)^2.
\end{equation}

Observe that although naive dimensional analysis
might suggest $C_{ij} \sim 1 / M_{pl}^5$, this is really a lower bound
on the strength of this interaction. The 5D Planck scale
is related to the 11D Planck scale $M_{11D}$ and the Kaluza-Klein scale $M_{KK}$ via:
\begin{equation}
\left(\frac{M_{KK}}{M_{11D}} \right)^6 \left(\frac{M_{pl}}{M_{11D}} \right)^3 \sim 1.
\end{equation}
Since we have assumed $L$ is large relative to the 11D Planck length anyway, this also means:
\begin{equation} \label{5Dhierarchies}
M_{pl}>M_{11D}>M_{KK}
\end{equation}
So returning to line (\ref{ccoeff}), we learn that $C_{ij} \gtrsim 1/M_{pl}^5$.
In other words the strength of the interaction between two CFTs is at
the very least dictated by the Planck scale, but there can be additional
enhancement when there is a hierarchy between the Kaluza-Klein scale and
the 5D Planck scale. This example also illustrates that although we can sometimes tune
the relevant deformations to zero (reaching a fixed point), irrelevant
deformations and interactions between different CFTs cannot be tuned away.


Let us now turn to 4D examples. Here, we consider stringy examples with ${\cal N}=4,2,1$ supersymmetries, and briefly comment on the non-supersymmetric case.

Consider first 4D $\mathcal{N} = 4$ theories. These SCFTs are labelled by a choice of semi-simple gauge group and
can all be viewed as descending from compactification of the 6D $(2,0)$ theories on a suitable $T^2$ (possibly
with twists). Coupling such sectors to supergravity has been studied in the literature (see e.g. \cite{Bergshoeff:1985ms}) but does
not appear to impose any significant constraint on the actual matter content of the theory.

By contrast, the stringy list of possibilities are quite limited. Much as in our discussion of
6D $(2,0)$ theories coupled to gravity and the bound of line (\ref{6Drankbound}), the
total rank is bounded above by:
\begin{equation}
r \leq 22.
\end{equation}
One way to get the maximal rank is by compactification of type II strings on $K3\times T^2$ or equivalently heterotic strings on $T^6$, leading to the Narain charge lattice $\Gamma_{6,22}$ with a gauge group of total rank 22.
The $6 \times 22$ deformations of the Narain lattice correspond to deforming the ${\cal N}=4$ matter sectors with the Coulomb branch VEVs in the Cartan of the group.  In this case there are no relevant deformations of ${\cal N}=4$ SCFTs preserving ${\cal N}=4$ SUSY.  There are also other compactifications known where we can also get reduced ranks, as in CHL strings \cite{Chaudhuri:1995fk} by
including automorphism twists of heterotic strings after compactification on torii.
Thus it is natural to conjecture that $r=22$ is the maximum rank of ${\cal N}=4$ supersymmetric theories that can be coupled to ${\cal N}=4$ supergravity in flat space.

Let us now proceed to 4D $\mathcal{N} = 2$ SCFTs coupled to $\mathcal{N} = 2$ supergravity.
There are a number of ways to generate examples of this sort.  The largest known class of examples is
obtained from compactification of type II strings on Calabi-Yau threefolds.
Let us in particular consider type IIB on Calabi-Yau threefolds.  The non-compact versions of these manifolds near singularities are known to lead in the IR to 4D ${\cal N}=2$  SCFTs \cite{Shapere:1999xr} (see also \cite{DelZotto:2011an, Xie:2015rpa, Chen:2016bzh}).
For example, if we consider hypersurface singularities of the type:
\begin{equation}
uv+f(z_1,z_2)=0
\end{equation}
where $u,v,z_1,z_2 \in \mathbb{C}$ and $f(z_1,z_2)$ is a quasihomogeneous polynomial,
we get an SCFT in the IR.  Let us assign weights $q_1,q_2$ to $z_1,z_2$ such that
\begin{equation}
f(\lambda^{q_1} z_1,\lambda^{q_2}z_2)=\lambda f(z_1,z_2).
\end{equation}
Deformations away from the fixed point correspond to adding monomials in $z_1^{a}z_2^b$ to $f(z_1,z_2)$.
Such a deformation has weight $a q_1 + b q_2$.
We can organize the various types of deformation away from the fixed point using the quantity $w= (1-q_1-q_2)$.
In particular, we have:
\begin{align}
\text{Operator VEVs}:\,\,\,& 0 \leq a q_1 + b q_2 < w \\
\text{Relevant Deformation}:\,\,\,& w \leq a q_1 + b q_2 < 1 \\
\text{Marginal Deformation}:\,\,\,& a q_1 + b q_2 = 1 \\
\text{Irrelevant Deformation}:\,\,\,& 1 < a q_1 + b q_2 \leq 2w.
\end{align}
There are, of course many additional deformations which are automatically set to zero in the deformation ring. These
are all irrelevant operator deformations of the SCFT.

To illustrate, consider the special case:
\begin{equation}
f(z_1,z_2)= z_1^n + z_2^n
\end{equation}
which is known as the $(A_{n-1},A_{n-1})$ Argyres-Douglas theory (see e.g. \cite{Cecotti:2010fi}).
In this case $q_1=q_2=1/n$ and monomials in the singularity deformation ring are given by $z_1^a z_2^b$ which
has weight $(a + b) / n$. The ring of deformations is generated by those monomials with
$0 \leq a,b \leq n - 2$. In the physical theory, these
sort according to the following inequalities:
\begin{align}
\text{Operator VEVs}:\,\,\,& 0 \leq a + b <  n - 2 \\
\text{Relevant Deformation}:\,\,\,& n - 2 \leq a + b < n \\
\text{Marginal Deformation}:\,\,\,& a + b = n \\
\text{Irrelevant Deformation}:\,\,\,& n < a + b \leq 2n-4.
\end{align}

We can now ask whether this and related examples of $\mathcal{N} = 2$ SCFTs can be coupled to quantum gravity.
From the perspective of string compactification, this is equivalent to asking whether we can find such singularities in
\textit{compact} Calabi-Yau threefolds.\footnote{Note that the non-renormalization theorems of type IIB strings imply that the complex structures, which are part of vector multiplets, do not get deformed by quantum corrections since the string coupling is in a hypermultiplet. In other words, these statements are exact quantum mechanically as well.} First of all, since it is widely believed that there are only a finite
number of Calabi-Yau threefolds, this would immediately imply only a finite number of such models can appear
in quantum gravity. Far more non-trivial is that this set is non-empty: some of these examples
\textit{consistently embed} in compact examples.


To illustrate, consider type IIB on a Calabi-Yau threefold given
by an elliptic fibration over a base $\mathbb{P}^1 \times \mathbb{P}^1$.
The minimal Weierstrass model for this geometry is:
\begin{equation}
y^2 = x^3 + x f_{8,8} + g_{12,12}
\end{equation}
where $f_{8,8}$ is homogeneous of bidegree $(8,8)$, i.e., it is a polynomial of degree
$8$ in homogeneous coordinates of each $\mathbb{P}^1$, and $g_{12,12}$
has bidegree $(12,12)$. Consider, in the affine patch near the origin the following choices:
\begin{align}
f_{8,8} &= -\frac{3}{4} +  \beta z_{1}^{8} z_{2}^{8} \label{f88}\\
g_{12,12} &= \frac{1}{4} + z_{1}^{12}+ z_{2}^{12}.
\end{align}
The higher order term in line (\ref{f88}) corresponds to an
irrelevant deformation of the local singularity structure near $z_1 = z_2 = 0$.
The local geometry after a
suitable coordinate shift in $x$ and $y$, results in the
$(A_{11}, A_{11})$ Argyres-Douglas theory:
\begin{equation}
uv + z_{1}^{12} + z_{2}^{12} = 0.
\end{equation}
As can be seen from this example in a quantum theory of gravity we can fine tune not only all the relevant operator VEVs and deformations but also some of the irrelevant operators that could appear. For example, keeping the higher order terms we see that
some of the irrelevant deformations are also set to zero and the first irrelevant term that appears in $f_{8,8}$ is $z_1^8z_2^8$.

Descending to theories with even less supersymmetry, there are several known constructions of 4D $\mathcal{N} = 1$ SCFTs in
limits where 4D gravity is decoupled. This includes compactifications of $(1,0)$ theories from 6D,
various singular limits in local geometries, as well as brane probes of singular geometries and intersecting branes.

Coupling to gravity is more challenging because in addition to specifying a background geometry, it is also necessary to include the effects of
fluxes and non-perturbative contributions to the low energy effective theory including possible generation of superpotentials. In a globally complete model the number of flux quanta and branes which can be introduced is also strongly constrained by tadpole cancellation considerations \cite{Sethi:1996es}. This again imposes a finiteness condition on possible SCFTs coupled to gravity. Finally, while it would of course be interesting to discuss stringy examples with completely broken supersymmetry in flat space, there are at present no completely controlled examples of this sort. We leave this topic for future work.

\subsection{Phenomenological Considerations}

Our main conjecture is that embedding an effective field theory
in quantum gravity only allows a finite number of tunings. From this perspective, it is natural
to ask about the potential consequences for particle physics and cosmology. Here, we discuss two such aspects, one connected with the possibility of having additional decoupled extra sectors, and also the extent to which any effective field theory in quantum gravity
can be fine tuned.

In the context of stringy particle physics models, there can be many extra sectors beyond the Standard Model. Such sectors provide
dark matter candidates, and have also been considered in the model building literature as a possible means to source supersymmetry breaking effects. Sometimes the dynamics of these extra sectors can lead to problematic effects such as large flavor changing neutral currents so it is also common to posit that potentially problematic higher dimension operators can be suppressed, that is, ``sequestered'' (see e.g. \cite{Schmaltz:2006qs}). Some aspects of sequestering in string theory have been studied in \cite{Kachru:2007xp, Acharya:2018deu}.  It was found in reference
\cite{Berg:2010ha} that there is often some mixing for different field theory sectors, even when placed in different warped throats. Our conjecture on the appearance of only a finite number of fine tunings means there is an irreducible amount of mixing between any two QFT sectors in flat space which will happen with at least the strengths given by naive expectations of
gravitational effects related to inverse powers of $M_{pl}$.  On the constructive side, we note that we can have partially sequestered sectors where relevant operators are tuned to zero.

This circle of ideas is also of relevance in the specific context of stringy cosmological quintessence models, particularly
as motivated by general proposed swampland constraints on dark energy \cite{Obied:2018sgi, Agrawal:2018own}.
As discussed there, to be consistent with observations, a quintessence field can couple strongly only to the dark sector, and it is natural to view it as part of the dark sector.
The present considerations allow for some level of sequestering, and suggest
a general avenue for exploring this class of dark matter / quintessence models
which are very weakly coupled to the visible sector.

Finally, the fact that we can only tune a finite number of parameters in our low energy effective field theory also has bearing on the
stability of the electroweak scale in the Standard Model, namely that the mass of the Higgs is far smaller than the Planck scale.
One may naturally ask if quantum gravity has any bearing on this question.  Naively it may appear that it does not; however there have already been suggestions that the small value of neutrino masses may be related to a refined version of the weak gravity conjecture \cite{Ooguri:2016pdq, Ibanez:2017kvh, Hamada:2017yji}. Similarly, it has been suggested that the resolution of the hierarchy problem may have a swampland explanation \cite{Hamada:2017yji, Ibanez:2017oqr}. Such a fine tuning can arise in a quantum theory of gravity, so at least the present conjectures are not inconsistent with large hierarchies of scale. Needless to say, it is reassuring that the conditions we are proposing here are perfectly compatible with our observed Universe!

\section*{Acknowledgments}

JJH thanks the Harvard theory group for hospitality during the initial stages of this work.
The work of JJH is supported by NSF CAREER grant PHY-1756996.
The research of CV is supported in part by the NSF grant
PHY-1719924 and by a grant from the Simons Foundation (602883, CV).


\vspace{-3mm}


\begin{thebibliography}{99}                                                                                               %

\bibitem{Acharya:2006zw}
  B.~S.~Acharya and M.~R.~Douglas,
  ``A Finite Landscape?,''
  [hep-th/0606212].

\bibitem{Ooguri:2006in}
  H.~Ooguri and C.~Vafa,
  ``On the Geometry of the String Landscape and the Swampland,''
  Nucl.\ Phys.\ B {\bf 766}, 21 (2007)
  [hep-th/0605264].

\bibitem{LPV}
  D.~L\"ust, E.~Palti and C.~Vafa,
  \textit{to appear}.

\bibitem{Louis:2015mka}
  J.~Louis and S.~L\"ust,
  ``Supersymmetric AdS$_{7}$ backgrounds in half-maximal supergravity and marginal operators of (1, 0) SCFTs,''
  JHEP {\bf 1510}, 120 (2015)
  [arXiv:1506.08040 [hep-th]].

\bibitem{Cordova:2016xhm}
  C.~Cordova, T.~T.~Dumitrescu and K.~Intriligator,
  ``Deformations of Superconformal Theories,''
  JHEP {\bf 1611}, 135 (2016)
  [arXiv:1602.01217 [hep-th]].

\bibitem{Witten:1995zh}
  E.~Witten,
  ``Some comments on string dynamics,''
  [hep-th/9507121].

\bibitem{Gaiotto:2014kfa}
  D.~Gaiotto, A.~Kapustin, N.~Seiberg and B.~Willett,
  ``Generalized Global Symmetries,''
  JHEP {\bf 1502}, 172 (2015)
  [arXiv:1412.5148 [hep-th]].

\bibitem{AlvarezGaume:1983ig}
  L.~Alvarez-Gaum\'{e} and E.~Witten,
  ``Gravitational Anomalies,''
  Nucl.\ Phys.\ B {\bf 234}, 269 (1984).

\bibitem{Aspinwall:1996mn}
  P.~S.~Aspinwall,
  ``K3 surfaces and string duality,''
  In *Yau, S.T. (ed.): Differential geometry inspired by string theory* 1-95
  [hep-th/9611137].

\bibitem{Kumar:2010ru}
  V.~Kumar, D.~R.~Morrison and W.~Taylor,
  ``Global aspects of the space of 6D $\mathcal{N} = 1$ supergravities,''
  JHEP {\bf 1011}, 118 (2010)
  [arXiv:1008.1062 [hep-th]].


\bibitem{Morrison:2012np}
  D.~R.~Morrison and W.~Taylor,
  ``Classifying bases for 6D F-theory models,''
  Central Eur.\ J.\ Phys.\  {\bf 10}, 1072 (2012)
  [arXiv:1201.1943 [hep-th]].


\bibitem{Morrison:2012js}
  D.~R.~Morrison and W.~Taylor,
  ``Toric bases for 6D F-theory models,''
  Fortsch.\ Phys.\  {\bf 60}, 1187 (2012)
  [arXiv:1204.0283 [hep-th]].

\bibitem{Taylor:2011wt}
  W.~Taylor,
  ``TASI Lectures on Supergravity and String Vacua in Various Dimensions,''
  [arXiv:1104.2051 [hep-th]].

\bibitem{Heckman:2013pva}
  J.~J.~Heckman, D.~R.~Morrison and C.~Vafa,
  ``On the Classification of 6D SCFTs and Generalized ADE Orbifolds,''
  JHEP {\bf 1405}, 028 (2014)
  Erratum: [JHEP {\bf 1506}, 017 (2015)]
  [arXiv:1312.5746 [hep-th]].

\bibitem{Heckman:2015bfa}
  J.~J.~Heckman, D.~R.~Morrison, T.~Rudelius and C.~Vafa,
  ``Atomic Classification of 6D SCFTs,''
  Fortsch.\ Phys.\  {\bf 63}, 468 (2015)
  [arXiv:1502.05405 [hep-th]].

\bibitem{Heckman:2018jxk}
  J.~J.~Heckman and T.~Rudelius,
  ``Top Down Approach to 6D SCFTs,''
  J.\ Phys.\ A {\bf 52}, no. 9, 093001 (2019)
  [arXiv:1805.06467 [hep-th]].


\bibitem{DelZotto:2014fia}
  M.~Del Zotto, J.~J.~Heckman, D.~R.~Morrison and D.~S.~Park,
  ``6D SCFTs and Gravity,''
  JHEP {\bf 1506}, 158 (2015)
  [arXiv:1412.6526 [hep-th]].

\bibitem{Anderson:2018heq}
  L.~B.~Anderson, A.~Grassi, J.~Gray and P.~K.~Oehlmann,
  ``F-theory on Quotient Threefolds with (2,0) Discrete Superconformal Matter,''
  JHEP {\bf 1806}, 098 (2018)
  [arXiv:1801.08658 [hep-th]].

\bibitem{Hayashi:2019fsa}
  H.~Hayashi, P.~Jefferson, H.~C.~Kim, K.~Ohmori and C.~Vafa,
  ``SCFTs, Holography, and Topological Strings,''
  [arXiv:1905.00116 [hep-th]].

\bibitem{Taylor:2012dr}
  W.~Taylor,
  ``On the Hodge structure of elliptically fibered Calabi-Yau threefolds,''
  JHEP {\bf 1208}, 032 (2012)
  [arXiv:1205.0952 [hep-th]].

\bibitem{Grassi}
  A.~Grassi,
  ``On Minimal Models of Elliptic Threefolds,''
  Math.\ Ann.\ {\bf 290}, 287-301 (1991).

\bibitem{Gross}
  M.~Gross,
  ``A Finiteness Theorem for Elliptic Calabi-Yau Threefolds,''
  Duke.\ Math.\ Jour.\ {\bf 74}, 271 (1994).

\bibitem{Candelas:1997eh}
  P.~Candelas, E.~Perevalov and G.~Rajesh,
  ``Toric geometry and enhanced gauge symmetry of F theory / heterotic vacua,''
  Nucl.\ Phys.\ B {\bf 507}, 445 (1997)
  [hep-th/9704097].

\bibitem{Aspinwall:1997ye}
  P.~S.~Aspinwall and D.~R.~Morrison,
  ``Point - like instantons on K3 orbifolds,''
  Nucl.\ Phys.\ B {\bf 503}, 533 (1997)
  [hep-th/9705104].

\bibitem{KSV}
  H.-C.~Kim, G. Shiu and C. Vafa,
   ``Branes and the Swampland,'' \textit{to appear}.

\bibitem{DelZotto:2017pti}
  M.~Del Zotto, J.~J.~Heckman and D.~R.~Morrison,
  ``6D SCFTs and Phases of 5D Theories,''
  JHEP {\bf 1709}, 147 (2017)
  [arXiv:1703.02981 [hep-th]].

\bibitem{Jefferson:2017ahm}
  P.~Jefferson, H.~C.~Kim, C.~Vafa and G.~Zafrir,
  ``Towards Classification of 5d SCFTs: Single Gauge Node,''
  [arXiv:1705.05836 [hep-th]].

\bibitem{Morrison:1996xf}
  D.~R.~Morrison and N.~Seiberg,
  ``Extremal transitions and five-dimensional supersymmetric field theories,''
  Nucl.\ Phys.\ B {\bf 483}, 229 (1997)
  [hep-th/9609070].

\bibitem{Douglas:1996xp}
  M.~R.~Douglas, S.~H.~Katz and C.~Vafa,
  Nucl.\ Phys.\ B {\bf 497}, 155 (1997)
  [hep-th/9609071].


\bibitem{Chaudhuri:1995fk}
  S.~Chaudhuri, G.~Hockney and J.~D.~Lykken,
  ``Maximally supersymmetric string theories in D < 10,''
  Phys.\ Rev.\ Lett.\  {\bf 75}, 2264 (1995)
  [hep-th/9505054].

\bibitem{Bergshoeff:1985ms}
  E.~Bergshoeff, I.~G.~Koh and E.~Sezgin,
  ``Coupling of Yang-Mills to $\mathcal{N} = 4$, $D=4$ Supergravity,''
  Phys.\ Lett.\ B {\bf 155}, 71 (1985)
  [Phys.\ Lett.\  {\bf 155B}, 71 (1985)].

\bibitem{Cecotti:2010fi}
  S.~Cecotti, A.~Neitzke and C.~Vafa,
  ``R-Twisting and 4d/2d Correspondences,''
  arXiv:1006.3435 [hep-th].

\bibitem{Shapere:1999xr}
  A.~D.~Shapere and C.~Vafa,
  ``BPS structure of Argyres-Douglas superconformal theories,''
  hep-th/9910182.

\bibitem{DelZotto:2011an}
  M.~Del Zotto,
  ``More Arnold's $\mathcal{N} = 2$ superconformal gauge theories,''
  JHEP {\bf 1111}, 115 (2011)
  [arXiv:1110.3826 [hep-th]].

\bibitem{Xie:2015rpa}
  D.~Xie and S.~T.~Yau,
  ``4d $\mathcal{N} = 2$ SCFT and singularity theory Part I: Classification,''
  [arXiv:1510.01324 [hep-th]].

\bibitem{Chen:2016bzh}
  B.~Chen, D.~Xie, S.~T.~Yau, S.~S.-T.~Yau and H.~Zuo,
  ``4D $\mathcal{N} = 2$ SCFT and singularity theory. Part II: complete intersection,''
  Adv.\ Theor.\ Math.\ Phys.\  {\bf 21}, 121 (2017)
  [arXiv:1604.07843 [hep-th]].

\bibitem{DelZotto:2015rca}
  M.~Del Zotto, C.~Vafa and D.~Xie,
  ``Geometric engineering, mirror symmetry and $ 6{\mathrm{d}}_{\left(1,0\right)}\to 4{\mathrm{d}}_{\left(\mathcal{N}=2\right)} $,''
  JHEP {\bf 1511}, 123 (2015)
  [arXiv:1504.08348 [hep-th]].


\bibitem{Sethi:1996es}
  S.~Sethi, C.~Vafa and E.~Witten,
  ``Constraints on low dimensional string compactifications,''
  Nucl.\ Phys.\ B {\bf 480}, 213 (1996)
  [hep-th/9606122].


\bibitem{Schmaltz:2006qs}
  M.~Schmaltz and R.~Sundrum,
  ``Conformal Sequestering Simplified,''
  JHEP {\bf 0611}, 011 (2006)
  [hep-th/0608051].

\bibitem{Kachru:2007xp}
  S.~Kachru, L.~McAllister and R.~Sundrum,
  ``Sequestering in String Theory,''
  JHEP {\bf 0710}, 013 (2007)
  [hep-th/0703105 [hep-th]].

\bibitem{Acharya:2018deu}
  B.~S.~Acharya, A.~Maharana and F.~Muia,
  ``Hidden Sectors in String Theory: Kinetic Mixings, Fifth Forces and Quintessence,''
  JHEP {\bf 1903}, 048 (2019)
  [arXiv:1811.10633 [hep-th]].

\bibitem{Berg:2010ha}
  M.~Berg, D.~Marsh, L.~McAllister and E.~Pajer,
  ``Sequestering in String Compactifications,''
  JHEP {\bf 1106}, 134 (2011)
  doi:10.1007/JHEP06(2011)134
  [arXiv:1012.1858 [hep-th]].

\bibitem{Obied:2018sgi}
  G.~Obied, H.~Ooguri, L.~Spodyneiko and C.~Vafa,
  ``De Sitter Space and the Swampland,''
  arXiv:1806.08362 [hep-th].

\bibitem{Agrawal:2018own}
  P.~Agrawal, G.~Obied, P.~J.~Steinhardt and C.~Vafa,
  ``On the Cosmological Implications of the String Swampland,''
  Phys.\ Lett.\ B {\bf 784}, 271 (2018)
  [arXiv:1806.09718 [hep-th]].

\bibitem{Ooguri:2016pdq}
  H.~Ooguri and C.~Vafa,
  ``Non-supersymmetric AdS and the Swampland,''
  Adv.\ Theor.\ Math.\ Phys.\  {\bf 21}, 1787 (2017)
  [arXiv:1610.01533 [hep-th]].

\bibitem{Heckman:2019dsj}
  J.~J.~Heckman, C.~Lawrie, L.~Lin, J.~Sakstein and G.~Zoccarato,
  ``Pixelated Dark Energy,''
  arXiv:1901.10489 [hep-th].

\bibitem{Heckman:2018mxl}
  J.~J.~Heckman, C.~Lawrie, L.~Lin and G.~Zoccarato,
  ``F-theory and Dark Energy,''
  arXiv:1811.01959 [hep-th].

\bibitem{Ibanez:2017kvh}
  L.~E.~Ib\'{a}\~{n}ez, V.~Martin-Lozano and I.~Valenzuela,
  ``Constraining Neutrino Masses, the Cosmological Constant and BSM Physics from the Weak Gravity Conjecture,''
  JHEP {\bf 1711}, 066 (2017)
  [arXiv:1706.05392 [hep-th]].

\bibitem{Hamada:2017yji}
  Y.~Hamada and G.~Shiu,
  ``Weak Gravity Conjecture, Multiple Point Principle and the Standard Model Landscape,''
  JHEP {\bf 1711}, 043 (2017)
  [arXiv:1707.06326 [hep-th]].

\bibitem{Ibanez:2017oqr}
  L.~E.~Ib\'{a}\~{n}ez, V.~Martin-Lozano and I.~Valenzuela,
  ``Constraining the EW Hierarchy from the Weak Gravity Conjecture,''
  [arXiv:1707.05811 [hep-th]].

\end{thebibliography}
\end{document}